# Exploiting ecological principles to better understand cancer progression and treatment

**David Basanta and Alexander R. A. Anderson**
Integrated Mathematical Oncology, Moffitt Cancer Centre, Tampa, FL, 33629, USA.
*Correspondence: David@CancerEvo.org and Alexander.Anderson@Moffitt.org

A small but growing number of people are finding interesting parallels between ecosystems as studied by ecologists (think of a Savanna or the Amazon rain forest or a Coral reef) and tumours[1-3]. The idea of viewing cancer from an ecological perspective has many implications but fundamentally, it means that we should not see cancer just as a group of mutated cells. A more useful definition of cancer is to consider it a disruption in the complex balance of many interacting cellular and microenvironmental elements in a specific organ. This perspective means that organs undergoing carcinogenesis should be seen as sophisticated ecosystems in homeostasis that cancer cells can disrupt. It also makes cancer seem even more complex but may ultimately provides isights that make it more treatable. Here we discuss how ecological principles can be used to better understand cancer progression and treatment, using several mathematical and computational models to illustrate our argument.

## 1. Cancer and ecosystems

One of the primary aims of mathematical modelling is to make the system being studied more understandable. This often means defining the system as simply as possible, and not making it more complex than reality. Einstein is known to have said that everything should be made as simple as possible, but not simpler[4]. It turns out that complexity has its place and, as convenient as it would be for cancer biologists to study tumour cells in isolation, that makes as much sense as trying to understand frogs without considering that they tend to live near swamps and feast on insects. A frog's sticky tongue makes much more sense when you consider how useful is it when trying to catch flies. Similarly, it makes sense that a cancer cell that is close to a blood vessel and is capable of producing Vascular Endothelial Growth Factors (VEGF) would benefit from co-opting endothelial cells to grow its very own vasculature and obtain more nutrients and oxygen. This dialogue between tumour cells and their environment is critical to understanding how an ecological view of cancer may be beneficial. The standard gene centric view states that cancer is only a product of mutation but since the importance of a mutation only makes sense when we understand the context. The context in which genes operate is ultimately the **ecosystem**. An ecosystem is made of individuals (plants, animals, bacteria, independent cells,...) and the physical environment they inhabit (water, soil, oxygen, food, etc). Survival and proliferation, the only things that matter at the evolutionary level, depend on how well a cell competes for the existing resources and cooperates with other cells to produce new ones. Even a simplified ecosystem should showcase the interdependence of species and how important the interactions between them are. In many ecosystems, the species and the way they interact do not change significantly over time. Occasionally, changes in the environment or a new species invading the ecosystem can disrupt the existing homeostasis. From the cancer ecology perspective, tumourigenesis is the process by which the homeostasis that characterises a healthy tissue is disrupted either via changes in the tissue microenvironment, or by an invading species (some bacteria and viruses are known to be able to lead to tumourigenesis), or a local invasion (a resident species producing a brand new one as a result of mutations).

The local environment is thus an important factor, not only in traditional ecosystems but also in cancer. This idea dates back to the late 19th century with Paget's well known seed-soil hypothesis [5] which suggests that for successful metastases, the soil (the site of a metastasis) is as important as the seed (the metastatic cells). It is now beginning to be widely accepted that cancer is not just a genetic disease but one in which **evolution** plays a crucial role[1,3]. This means that tumour cells evolve, adapt to and change the environment in which they live. The ones that fail to do so will ultimately become extinct. The ones that do, will have a chance to invade and metastasise. The capacity of a tumour cell to adapt to a new environment will thus be determined by environment and the cellular species from the original site, to which it has already painstakingly adapted.

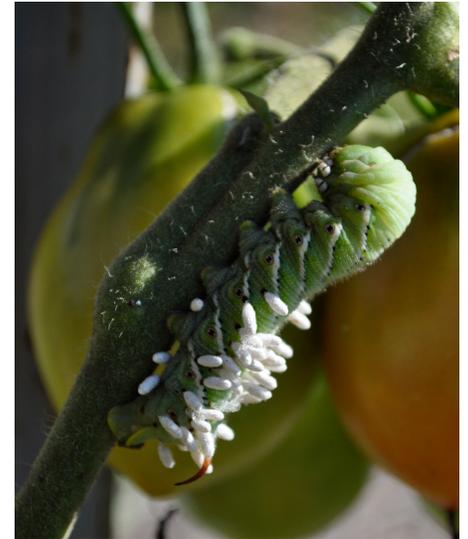

Figure 1. Parasitism: Parasitic wasp cocoons attached to a caterpillar. By Jacob Scott, M.D.

**Adaptation** is the key process for any system subject to Darwinian evolution, and cancer is no exception [6]. Because its role in Cancer is only now beginning to be explored, the full implications have yet to impact the cancer research community at large. Tumour cell adaptation to complex dynamic environments not only means that finding the roots of the disease just got a whole lot more complex (as it is not restricted to the role of a few genes) but this view also opens new routes to stop or even reverse cancer progression [7]. Normal organ ecosystems (and the tissues that define them) maintain a dynamic balance or homeostasis between their cellular and environmental components and therefore do not create selection pressures that lead to adaptation.

This **homeostasis** is a defining feature of normal healthy body organs (such as those in which cancer has not been initiated). Evolution selects for organisms that achieve homeostasis and this evolutionary process also makes them capable of recovering from environmental and genetic perturbations [8,9]. The normal form and function of most tissues (defined by the integration of multiple cellular, extracellular, chemical and physical signals/constraints) is to maintain a homeostatic balance and carry out the role they are required to perform. Homeostasis loss is traditionally seen as a key initial step on the route to cancer development [10,11]. At its simplest tissue homeostasis is the balance between cell proliferation and apoptosis such that the tissue architecture and function remains constant. It is no accident that disruptions in these processes are considered as key features of oncogenic transformation. Fortunately, there are multiple mechanisms that regulate these processes and actively ensure homeostatic maintenance, mainly through the regulation of both proliferation and apoptosis. These mechanisms fall into the two broad camps of cellular (e. g. cell-cell adhesion, cell-ECM adhesion), and environmental (e.g. metabolic factors, growth factors, stroma) although there is a great deal of feedback between these camps with changes in one driving the other. Therefore to escape homeostatic control mutant cells need to significantly modify their baseline phenotypes and ignore environmental signals. This will be profoundly influenced by both cellular (in terms of phenotypic traits such as cell adhesion) and environmental heterogeneity (in terms of metabolite levels and stromal communication) and the feedback between them.

**Cellular heterogeneity** represents an intrinsic variability that may be driven by genetic or non-genetic factors but importantly provides the means for driving homeostatic disruption and responding to it. This heterogeneity further emphasises the need to understand interactions that occur within the cancer ecosystem i.e. between cells and between cells and their environment. Genetic heterogeneity within a tumour is often referred to as intratumour heterogeneity and is currently of great interest to the cancer community as it has highlighted a potential issue with molecular signatures and even personalised medicine as it is currently understood. Specifically, Swanton and colleagues[12] have shown that multiple biopsies from the same tumour display distinct genetic profiles and yet are phenotypically similar. This genotypic divergence and phenotypic convergence has also been observed across many different cancers, including those originating in the lung, Kidney [12], prostate and brain [13,14]. We believe that this disparity, between genotype and phenotype, is a natural result of the organ ecosystems that these cancer originate in, regardless of the specific mutations that may facilitate the cancers progression the intricate dialogue between the tumour cells and their environment selects for clones that are best adapted phenotypically to survive.

From our discussion above it is clear to us that tumours are made of a heterogeneous mixture of cells and that tumour heterogeneity, which manifests itself at the genotypic and phenotypic level, affects the way that tumour cells interact with other tumour cells, other healthy cells as well as the physical microenvironment. Furthermore, these interactions can drive the behaviour of healthy cells such as fibroblasts, which can be permanently transformed into carcinoma associated fibroblasts (CAFs)[40] under the right conditions. An ecosystem view of cancer dictates that cancer progression is a complex process that emerges from the interactions between individuals and their environment. How can we hope to understand tissue homeostasis and evolution-driven disruption that leads to cancer? Purely experimental approaches are unpractical given the complexity

---

**Box 1: The Hawk-Dove game**

Imagine a species in which individuals could have either aggressive or meek strategies to resolve disputes over food. Let's call the former hawks and the latter doves. When two doves have to share food (which we will refer as V) they just divide it in two halves (each getting V/2). When two hawks dispute over food they fight and the victor takes the spoils (V) whereas the loser is assumed to be severely harmed (-C). In the third scenario, when a hawk and a dove meet, the dove balks away from the fight leaving all the food to the hawk. This information is captured in the payoff table shown below. EGT can tell us a few things about this population. For instance, the obvious one: that a population made of dove-like individuals is susceptible to be invaded by a few hawks. Intriguingly we can also learn that in many cases a population made of hawks is unlikely to be immune to invasion by a handful of doves: assuming that hawks fight until one of them is severely beaten then a dove that doesn't fight might go hungry before it meets another dove but that still beats being severely injured. With a little information about how serious the average injury would be and how much a given resource would help reproduction we could guess what the proportion of aggressive versus meek individuals would be in the long term. This is what, in game theoretical parlance, is known as an evolutionary stable set of strategies, which implies that the ecosystem is at an equilibrium that will not be easily disrupted. A population in this type of equilibrium will recover from perturbations, even if part of the population changes strategy (unless, of course, we consider alternative phenotypic strategies to

|  | Hawk | Dove |
|---|---|---|
| Hawk | $\frac{V-C}{2}$ | $V$ |
| Dove | $0$ | $\frac{V}{2}$ |

hawk and dove). Similar games can be played with tumour cell populations. A good example would be a tumour with tumour phenotypes that move away when confronted with scarce resources (motile) and tumour phenotypes that stay to use them (proliferative). This scenario was studied using a Hawk-Dove game[15]. In our view, a multicellular organism can be seen as a group of cells in a dynamic equilibrium (known as homeostasis) that is robust and stable for phenotypic strategies that are normal in a healthy organism but not necessarily to cancerous ones. This phenotypic approach to study cancer can be quite useful. It is widely accepted that tumour cells acquire a number of new phenotypical capabilities on the path towards malignancy[10]. The interactions among different tumour phenotypes can be studied using EGT to investigate the possible sequence of steps that characterise cancer progression as well as the circumstances that lead to the emergence of increasingly aggressive phenotypes.

of interactions and time scales involved in cancer. Fortunately there already exists mathematical and computational tools that can be used to study ecosystems regardless of size, scale, and complexity.

## 2. Mathematical tools

One good example of a mathematical tool to study evolution in ecosystems is game theory (GT). GT was initially introduced to understand human and sociological behaviour. With GT we can study games in which the outcome affecting a player depends, not only on the strategy used, but on the strategies employed by the other players. A key aspect is that a game strategy is not good or bad considered in isolation. Only when compared with the strategies employed by other players can we make that call. John Maynard Smith pioneered the use of this tool to study evolutionary dynamics in biology. This is known as evolutionary game theory (**EGT**). The GT assumption that players have to be rational is, paradoxically, better suited to the individuals in sociology, economics or war. The force of natural selection keeps ecosystem denizens focused on optimising the bottom line: long term reproduction. In the games studied by evolutionary game theoreticians, individuals compete for available resources using a variety of strategies. These features and behaviours, known as the phenotypic strategy, determine the winners and losers of evolution. We, and others, have developed simple mathematical models using EGT to understand homeostasis and its disruption in cancer (for an example of an evolutionary game take a look at box 1). Many lessons can be learned from very simple games like that one.

1. One crucial lesson, especially when used to understand cancer evolution, is that <u>focusing on indiscriminately destroying as many cancer cells as possible is not necessarily the best thing to do for a patient</u>. In EGT, the long term (equilibrium) outcome of a game depends on the interactions between the players, not on the size of the population. A treatment based exclusively on indiscriminately removing most (but not all) cancer cells may have only a temporary effect as in most cases the original number of tumour cells will eventually be restored and exceeded. Many EGT models show that a more effective alternative would be based on changing the way cells

|     | AG              | INV             | GLY                       |
|-----|-----------------|-----------------|---------------------------|
| AG  | $\frac{1}{2}$   | $1-c$           | $\frac{1}{2}+n-k$         |
| INV | $1$             | $1-\frac{c}{2}$ | $1-k$                     |
| GLY | $\frac{1}{2}-n$ | $1-c$           | $\frac{1}{2}-k$           |

The three phenotypes in the game are autonomous growth (AG), invasive (INV) and glycolytic (GLY). The base payoff in a given interaction is equal to 1 and the cost of moving to another location with respect to the base payoff is $c$. The fitness cost of acidity is $n$ and $k$ is the fitness cost of having a less efficient glycolytic metabolism. The table should be read following the columns, and thus the fitness change for an invasive cell interacting with an autonomous growth would be $1-c$.

   interact with each other and their environment which would affect their fitness and thus, potentially, drive tumour evolution towards less aggressive cell types or at least to a stable coexistence that would be less harmful to the patient [7].

2. Intra tumour <u>heterogeneity</u> is a crucial property of cancer, and the dynamics of a tumour can change dramatically if more phenotypes with different traits emerge in a population of cancer cells. A good example of this can be seen in [15,16], a variation of the game explained in box 1. In this version of the game we consider the two phenotypes of the original game: invasive (INV) and proliferative (AG), and add a new one: glycolytic (GLY). The resulting table can be seen in the payoff table above this paragraph. In the original situation with two phenotypes, the presence of more aggressive motile phenotypes depended almost exclusively on the cost of motility (cost:an abstraction for the fitness that moving cells pay in terms of degrading the extra cellular matrix and avoiding anoikis). The addition of a new phenotype alters the game by changing the incentives in favour of motile phenotypes. By making proliferative phenotypes less successful, glycolytic

|   | S | D | I |
|---|---|---|---|
| S | 0 | $\alpha$ | 0 |
| D | $1+\alpha-\beta$ | $1-2\beta$ | $1-\beta+\rho$ |
| I | $1-\gamma$ | $1-\gamma$ | $1-\gamma$ |

The fitness of each of the phenotypes (S, Stroma; D, microenvironmentally dependent; I, microenvironmentally independent) depends on the interactions with other phenotypes and the values of the costs and benefits resulting from these interactions. These costs and benefits are: $\alpha$ (benefit derived from the cooperation between a S cell and a D cell), $\gamma$ (cost of being microenvironmentally independent), $\beta$ (cost of extracting resources from the microenvironment and $\rho$ (benefit derived by D from paracrine growth factors produced by I cells.

phenotypes increase the relative fitness advantage of motile phenotypes resulting in more aggressive tumours than in the original situation. In general EGT can be used to explore how changes in tumour's phenotypic heterogeneity could change the evolutionary dynamics in a cancer even if, as in the example, that intermediate phenotype might not be evolutionarily successful in the long term.

3. Modern clinical cancer research is betting that personalised treatments and targeted therapies are our best shot at providing durable cures to many types of cancer. Understanding the impact of targeted treatments in cancer is easier with tools like EGT where the effects of selective therapies in heterogeneous tumours can be studied. Using a model described in the above table [17] we have explored how treatments could affect tumour heterogeneity and clonal dominance. An example from this game can be seen in figure 2 showing a tumour with two main clonal populations (D, I), and a stromal population (S) that has been co-opted to help the more successful tumour population (D, in this scenario). In this case, treatment designed to impact the maximum possible number of tumour cells has left behind a smaller tumour population and its growth potential unaffected. Furthermore, the resulting tumour, which initially was incredibly susceptible, is now completely resistant to the treatment.

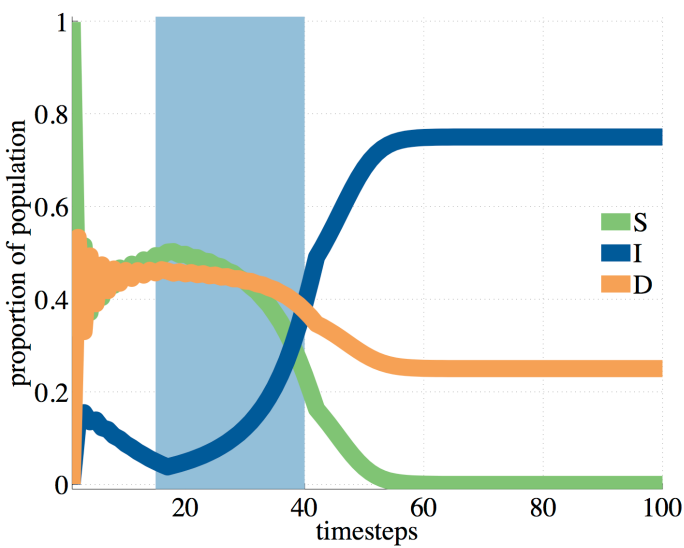

Figure 2. Result from a replicator equation resulting from a game where certain stromal cells (S) and certain tumour cells (D) cooperate. Treatment (sky blue) kills stromal cells effectively selecting for the I tumour population.

We will produce better treatments if we use evolution in our favour instead of ignoring it as the driving force of tumour progression. An ecologically enlightened approach would take into account what we learned before: that killing the most tumour cells might not be the best strategy, that intra-tumour heterogeneity increases the chances of tumour recurrence after treatment and that treatments represent a form of selection - where phenotypes that are not selected against will survive and lead to not only recurrent cancer but resistant cancer. An approach where tumour diversity could be reduced in stages, and treatments chosen so they select for increasingly more benign or easier to treat tumours. An example of such an approach was initially proposed by Merlo and colleagues [1] who dubbed it *sucker's gambit*. The idea of sucker's gambit is that we could change the selection pressure in an evolving tumour such that the easier to treat phenotypes would be selected for. Subsequently, Gatenby, Brown and Vincent suggested using insights gained in ecology to further explore the idea of the "sucker's gambit". With an evolutionary double bind, the species whose numbers need to be controlled must be predated by at least two different types of predators [18,19]. The secret of a successful evolutionary double bind is that the strategies of the two predator species have to be synergistic such that a prey evolving in order to be attacked by one of the predators will in fact become more susceptible to being attacked by the other. This type of strategy is easy to conceive as a EGT model and recently we have framed clinical results from Antonia and colleagues [20,21] in an evolutionary double bind to try to explain their results. In their research, Antonia and colleagues show that when applying two different therapies to lung cancer patients, a p53 vaccine and standard chemotherapy, the order of the application of treatments has a substantial impact on the efficacy of the overall treatment. Specifically, patients that went through chemotherapy before the p53 vaccine was administered responded much better than patients where only one treatment was used or those where the sequence was reversed. The EGT model considered three populations: those susceptible to all treatments, those resistant to chemotherapy and those resistant to the p53 vaccine [22]. The model highlights the importance of finding the right sequence of treatments such that evolution can be directed to tumours that are easier to treat.

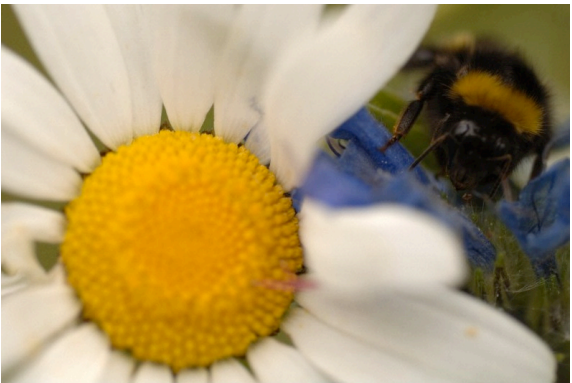

Figure 3. Symbiosis. Both the bee and the flower derive a benefit from their interaction. Wellcome Library, London

**Interactions and cooperation**

Another potential use of the cancer-ecosystem viewpoint is the study of the evolutionary dynamics leading to the emergence of **cooperation** [2,23,24]. A common misunderstanding about evolution is that the *survival of the fittest* means that only the strongest and meanest survive and that they have done so through competition only. But nature is abundant with examples of inter and intra species cooperation. The trick is that cooperation can only emerge within the constraints of selection: it can only be sustainable if everybody, or at least the genes that promote cooperation, benefits from their interactions.

**Interactions** between individuals (people, animals, cells, bacteria) are normally categorised as mutualistic (figure 3), competitive, predatory, parasitic (figure 1) and comensalistic (see box 2)[25]. EGT is particularly useful at studying the interactions between the

> **Box 2: Competition and Cooperation**
>
> Mutualistic interactions are those in which both interacting parties benefit equally. Competitive interactions are those in which both parties have a detrimental effect on each other. Predatory and parasitic interactions are similar in that one of the parties benefits from the interactions whereas the other suffers, but in the former the harmed party will significantly reduce its fitness and in the latter the fitness reduction is much smaller. Finally comensalism describes interactions in which one of the parties derives neither harm nor gain whereas the other obtains some (often small) benefit. This table shows the costs (-) and benefits (+) of interactions between two parties (A and B).
>
> | A | B | Type |
> |---|---|------|
> |   |   | Neutralism |
> | - |   | Amensalism |
> | + |   | Comensalism |
> | - | - | Competition |
> | + | + | Mutualism |
> | + | - | Predation/parasitism |
>
> All these interactions may occur, to some degree or the other, in a tumour ecosystem. For instance tumour cells compete for space and nutrients but, intriguingly they may also team up to produce enough growth factors to sustain tumour growth[23].

players, how those affect tumour Darwinian evolution, and how evolution might lead to or away from homeostasis. However there are other important aspects in an ecosystem that can be approached in more detail with the help of other mathematical tools.

Most EGT models do not explicitly capture the role of space in an ecosystem, thus assuming that players interact with each other with a frequency that is only dependent on the relative proportion of each subpopulation. EGT models are also usually concerned with the relative proportion of the different subpopulations but, given the limitations of the approach, rarely do they study the absolute numbers in the population. Absolute numbers could be inferred from a properly parameterised model if the initial subpopulations are known as the fitness payoffs represent long-term proliferation rates. As it has already been reported before, many EGT models use more abstract parameters so that meaningful extrapolations between fitness and population size are not possible.

These limitations are not as severe as they might seem: many as of yet unaddressed questions can be answered by relatively more abstract approaches such as EGT. Its simplicity and focus on interactions makes it easy to understand the role of different subpopulations in a heterogeneous tumour. Alternatively, modelling tools in ecology can also use parameters derived from observations in the field. In cancer research there is abundant data resulting from *in vivo* and *in vitro* experiments at both the molecular and cellular levels that could be integrated with the right mathematical tool. One good example of such an approach is individual based models where each cell in a tumour is given its own identity and where the properties of the tumour as a whole can naturally emerge from the interactions between the different cells and between the cells and their environment. **Individual based models** (IBM) are a large class of models that consider both space and time explicitly and offer and ideal methodology to integrate some features of the other approaches discussed here. One of the pioneering IBM approaches is cellular automata (CA), which as in the case of GT where first introduced by Von Neumann, and were first used to explore biological questions [26]. Theoretical ecologists have utilised CA to investigate population dynamics [27,28] and the role of space in the interactions between individuals [29]. Specifically they can incorporate detailed descriptions of the individual (tumour cell, fish, fox etc) defining its behaviour (migrate, reproduce, die etc) in a given context (Savannah, lake, muscle tissue etc). IBMs therefore capture the spatial and temporal variation that characterises real ecosystems allowing us to explore the robustness of key homeostatic mechanisms [8,9]. Moreover, they have been extensively used by the modelling community to look at many different biological systems focussing on how individuals and their interactions collectively drive different evolutionary outcomes.

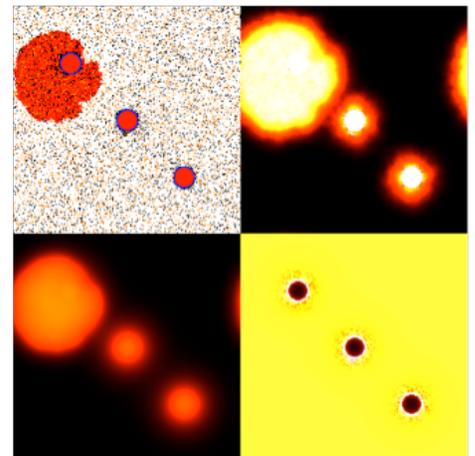

An IBM approach that has been used extensively in cancer modelling are **Hybrid models** that integrate both continuous and discrete variables and are able to incorporate biological phenomena on various temporal and spatial scales (See [30] for a recent review). These models represent cells as individual discrete entities and often use continuous concentration or density fields to model cell intracellular and extracellular environments. By their very nature, hybrid models are ideal for examining direct interactions between individual cells and between the cells and their microenvironment, but they also allow us to analyse the emergent properties of complex multicellular systems (such as cancer). It is worth noting that as these interactions take place on the intracellular and intercellular levels, but are manifested by changes on the tissue level, the emergent behaviour of growing multiclonal tumours are almost impossible to infer intuitively. Hybrid models can facilitate our understanding of the underlying biophysical processes in tumour growth. For example, by using high-throughput simulation techniques, we can examine the impact that changes in specific cell interactions (or their microenvironment) have on tumour growth and treatment. Hybrid models are often multiscale by definition integrating processes on different temporal and spatial scales, such as gene expression, intracellular pathways, intercellular signalling, cell growth, or migration. We have developed many hybrid models to investigate different aspects of cancer. For example, the Hybrid Discrete-continuum Cellular Automaton (HDC) model has been used to study how the interactions between tumour cells and stromal cells via a molecule known as TGFβ to explain prostate cancer progression. Figure 4 shows an example of the simulations produced by the model. With this approach it is possible to parameterise each cell independently using data collected from in vivo and in vitro experiments. Importantly, the mathematical model integrates all these biological data in a way that can yield clinically relevant insights by studying the emergent properties of the prostate cancer through timescales that typically cover decades.

Figure 4. Example of simulation in which an IBM is used to explore how the interactions between tumour cells and their environment affect progression.The screen on the top left shows tumour (red) and stromal (black and brown) interacting. Other screens show concentrations of elements of the physical microenvironment (TGFβ, Matrix Degrading Enzymes and Extra Cellular Matrix). An IBM model can shed light on the spatial distribution of relevant cellular species.

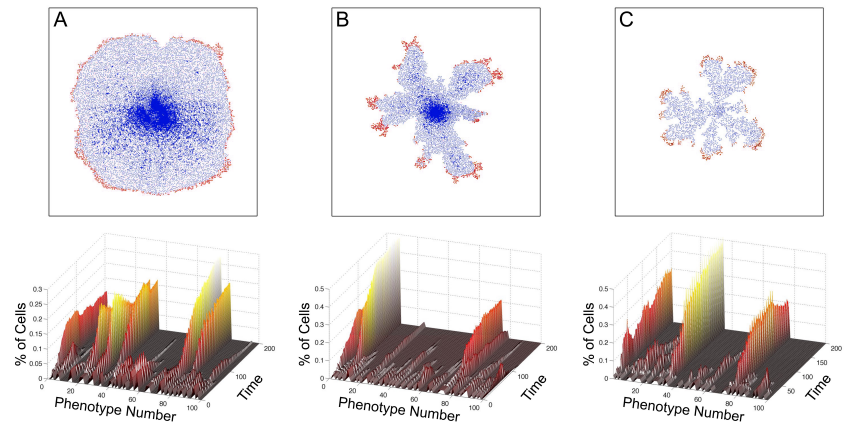

Figure 5. Simulation results from the HDC model under three different microenvironments: (A) uniform ECM, (B) Grainy ECM and (C) Low nutrient. The Upper row shows the resulting tumor cell distributions obtained after 3 months of simulated growth, we can see that the three different microenvironments have produced distinct tumour morphologies. The lower row shows the relative abundance of a possible 100 tumour phenotypes over time as the tumour invaded each of the different microenvironments. We note that there are approximately 6 dominant phenotypes in the uniform tumour, 2 in the grainy and 3 in the low nutrient tumour. These phenotypes have several traits in common: low cell-cell adhesion, short proliferation age, and high migration coefficients. In each tumour, one of the phenotypes is the most aggressive and also the most abundant, particularly in B and C. All parameters used in the simulations are identical with the exception of the different microenvironments.

Another major advantage of Hybrid models is their ability to easily incorporate heterogeneity both in terms of the tumour cell phenotype and the tumour microenvironment. Since interactions between tumours and their microenvironment drive selection and ultimately define the ecology of the tissue in which the tumour is developing, these models represent ideal tools to investigate evolution and selection in a growing tumour. Anderson and colleagues at Moffitt and Vanderbilt have shown (see figure 5) that different microenvironments (in terms of Extra Cellular Matrix density or nutrient concentration) will produce tumours with distinct morphological characteristics [6,31,32]. The research also shows that harsher microenvironments will select for more aggressive phenotypes (those that would lead to a worse prognosis for the patient) whereas *nicer* microenvironments could yield more heterogeneous tumours (where less aggressive clones coexist with more aggressive ones). These types of insights would be difficult to produce with only experimental and clinical data. Further research by the same group lead to a carefully parameterised version of the HDC model [33]. Using measurements of different cancer cell lines typically used in research labs across the world, the new model allowed them to study the effect of seeding *in vitro* different combinations of phenotypes in a number of micro-environments and made the surprising prediction that aggressive tumour cells have evolved to become essentially microenvironmentally independent (producing their own niche as required).

An extra strength of IBMs is that they can be combined with other mathematical tools like networks and EGT in order to complement each other's strength. As an example of the latter, Basanta and colleagues explored the emergence of motility in a tumour made of, essentially, proliferative cells using both EGT (see Box 1) and a CA-based IBM [34]. While the EGT implementation focused on the analysis of role of the cell-cell interactions in the evolutionary timescale (steady state), the CA provided insights on the role of space. Networks and IBM can also be combined, as shown by Gerlee and colleagues [35], so as to bridge the scales between the cellular level (the IBM) and the pathway level (the network). This allows for a model in which the behaviour of a cell is controlled by pathways (network nodes or vertices) and their interactions (network edges) as well as by the interactions with nearby cells and their microenvironment (CA lattice site).

## 3. Discussion

Traditionally the ecological perspective is firmly grounded at the scale of the phenotype and essentially ignores anything below this scale. It tends to be more encompassing at that scale and embraces all the different players of the ecosystem. In contrast with this perspective, the cancer biology view is very much centred on the genetic and molecular scales for which there is a wealth of data. Whilst this provides a solid foundation to work from for cancer ecologists, this data is unbalanced due to the poorly quantified phenotypic-scale. This imbalance is the result of the dominance and success of reductionism in cancer research. **Reductionism** is undoubtedly responsible for the exquisite level of understanding of many of the genes and pathways that are involved in tumour initiation and progression in a variety of organs. Both of these approaches have limitations but also have their own strengths that in fact compliment one another. Ideally we want to unify this biological-gene-centric view with the ecological-phenotype-centric view, however, experimentally this is difficult if not impossible, without the aid of theoretical approaches like the ones discussed above. In fact, there already exists IBMs that explicitly try to bridge the genotypic and phenotypic scales by incorporating elements of EGT and network theory [36,37].

The ecosystem view is, ultimately, a **holistic** one that sees cancer progression as a process that emerges from the interactions between multiple cellular species and interactions with the tumour microenvironment. An ecosystem perspective presents us with intriguing implications. One is that cancer is an evolutionary driven escape from homeostasis. It also casts aspects of cancer progression under a different light: are metastatic cells the ones that represent the best and most adapted cells at the primary site? Or, on the contrary, does metastasis and invasion represent the only alternative for less successful phenotypes, capable of escaping the primary site but unable to compete with better adapted ones locally? Could it be only a by-product of tumour cells acquiring the abilities to move and detach from the main body of the tumour? Is it the result of cooperation or competition? Regardless of the answer to these questions, an ecological interpretation of cancer would predict that metastasis will occur at sites in which the tumour cells will have a better chance of survival and colonisation. This will depend not only on the distance from the primary site or on the availability of lymphatic or blood vessels for physical connectivity [38] but also on the suitability of the new site for colonisation. Metastatic cells are already likely to be reasonably adapted to specific environmental conditions. A secondary site that somewhat resembles key features of the primary one while providing the metastatic cells with nutrients and room for growth will always be a more likely target for a secondary tumour.

One might ask, what would a cancer ecosystem look like? Unsurprisingly it will contain tumour cells, epithelial cells, nutrients and growth factors. Less intuitively it will also include immune and endothelial cells, nerves, different stromal phenotypes as well as epithelial cells in carcinomas. Figure 6 shows one example of an ecosystem for bone metastasis, though a simplification it already gives information in a purely visual way about the different types of interactions that are occurring. By adding weights to the interactions we can use the theoretical tools discussed above to investigate how a new tumour disrupts homeostasis (initiation), develops (growth), responds to perturbations

(treatment), evolves over time (progression) and how it may best be controlled or destroyed. The key point is that the ecosystem perspective places the emphasis on **interactions** and their consequences. A better understanding of these interactions could be used to hinder and even potentially reverse tumour progression. Tissue homeostasis disruption due to alterations in the tissue ecosystem could, potentially, be reversed via renormalisation of the tumour microenvironment [39]. For instance, it is known that normal stromal cells can inhibit progression towards malignancy in certain carcinomas [40]. Regaining homeostasis might not mean tumour eradication but instead may represent a new state where we live with cancer more like a chronic disease, kept in check by a combination of drugs that change in response to changes in the tumour or its microenvironment.

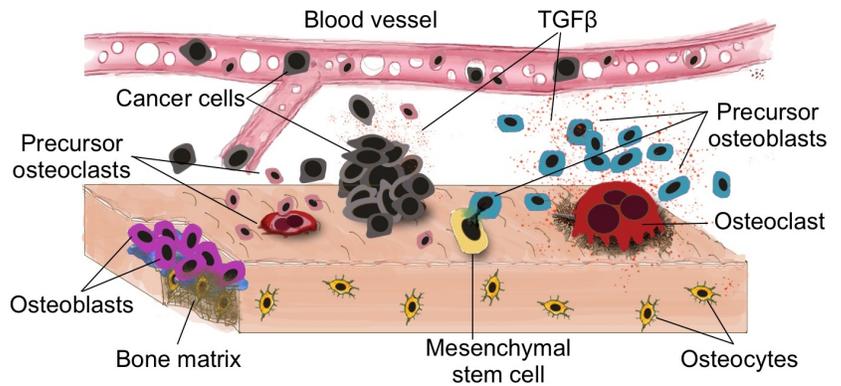

Figure 6. Outline of the ecosystem in a prostate to bone metastasis with several types of cancer cells interacting with other cellular populations like osteoblasts, osteoclasts, osteocytes and stem cells. Tumour cells interact to compete and cooperate for resources like nutrients, space and growth factors.

The timing for an ecosystemic view of cancer could not be better: with the development of high throughput automated microscopy the ability to gather substantial amounts of **cellular information** is becoming a reality. With this new information the cancer ecosystem is becoming more complete and therefore theoretical oncologists will have a better understanding of the key phenotypic strategies and mechanisms of interaction that tumour cells, and other relevant cells employ. Clearly this means we are more likely to be successful at producing models that are both holistic (taking into account the multiple scales at which cancer takes place) and quantitative (in which model parameters and predictions can be compared with experiments) i.e. qolistic approaches [41].

The heart of the matter is that an ecological view of tumours does not invalidate but **complements** and builds upon decades of cancer research and undoubtedly this will lead to a better understanding of the biology of cancer and to new and improved therapies. If we may use the old analogy but framed slightly differently: we need to properly understand the trees (e.g. every leaf, twig and branch) before we can understand the forest but we cannot afford to ignore the forest because the trees are so interesting on their own.

# Acknowledgements

We would to thank Arturo Araujo for the help with the figures in this paper. The work in this grant has been partly funded by DoD 12-16803-99-01, NIH U01 10-16381-01-01 and NIH U54 10-15885-03-03.